\documentclass[prd,aps,twocolumn,a4paper,showkeys,nofootinbib]{revtex4-1}

\usepackage{graphicx,psfrag}
\usepackage{mathrsfs}
\usepackage{amsmath,amsfonts,amssymb}
\usepackage[dvipsnames, usenames]{xcolor}
\usepackage{multirow}
\usepackage{comment}
\usepackage{multirow}
\usepackage{hyperref}

\def\d{{\rm d}}

\def\rhosat{\rho_{\rm sat}}
\def\Mo{{{\rm M}_\odot}}
\def\R14{{R_{1.4\Mo}}}

\def\pr{{\rm p}}

\def\data{\boldsymbol{d}}
\def\d{{\rm d}}
\def\kt{\kappa^{\rm T}_2}

\def\oldmodel{{\tt NRPM}}
\def\teob{{\tt TEOBResumS}}

\begin{document}

\title{
  Pre/post-merger consistency test for gravitational signals\\
  from binary neutron star mergers
}
	
\author{Matteo \surname{Breschi}$^{1,2,3}$} 
\author{Gregorio \surname{Carullo}$^{1,4}$} \author{Sebastiano \surname{Bernuzzi}$^{1}$} \affiliation{${}^{1}$Theoretisch-Physikalisches Institut,
	Friedrich-Schiller-Universit{\"a}t Jena, 07743 Jena, Germany}
\affiliation{${}^{2}$International School for Advanced Studies (SISSA), 34136 Trieste, Italy}
\affiliation{${}^{3}$INFN Sezione di Trieste, 34149 Trieste, Italy}
\affiliation{${}^{4}$Niels Bohr International Academy, Niels Bohr Institute, 2100 Copenhagen, Denmark}

\date{\today}

\begin{abstract}
  Gravitational waves from binary neutron star (BNS) mergers can
  constrain nuclear matter models predicting the neutron star's equation of
  state (EOS).
  Matter effects on the inspiral-merger signal are encoded in the
  multipolar tidal polarizability parameters,
  whose leading order combination is sufficient to capture to high
  accuracy the key features of the merger waveform 
  (e.g.~the merger frequency). 
  Similar EOS-insensitive relations exist for 
  the post-merger signal and can be used to model the emission from the remnant.
  Several works suggested that the appearance of new degrees of freedom
  or phase transitions in high-density post-merger matter can be
  inferred by observing a violation of these EOS-insensitive relations.
  Here, we demonstrate a Bayesian method to test such an EOS-insensitive relation between the tidal
  polarizability parameters (or any other equivalent parameter) and the dominant post-merger frequency,
  using information either up to merger or from the post-merger signal.
  Technically, the method is similar to tests of General Relativity with binary
  black holes that verify the inspiral-merger-ringdown consistency.
  However, differently from the latter, BNS pre/post-merger
  consistency tests are conceptually less informative and they  
  only address the consistency (or the breaking) of the assumed
  EOS-insensitive relation. 
  Specifically, we discuss how such tests cannot conclusively discriminate between an EOS not respecting such relation and the appearance of new degrees of freedom
  (or phase transitions) in high-density matter.
\end{abstract}

\pacs{
	04.25.D-,     
	04.30.Db,   
	95.30.Sf,     
	95.30.Lz,   
	97.60.Jd      
}

\maketitle

\section{Introduction}

Kilohertz gravitational waves (GWs) from binary neutron star (BNS) mergers
remnants are considered a
promising probe of the nuclear equation of state (EOS) at extreme density. 
While no such detection was possible for GW170817 \cite{Abbott:2017dke,Abbott:2018wiz,Abbott:2018hgk}, future experiments
are expected to reach the necessary sensitivity for a detection,
e.g.~\cite{Chatziioannou:2017ixj,Torres-Rivas:2018svp,Breschi:2021xrx}. 
Several authors claimed that
a viable path to constrain the extreme-densities EOS is to ``observe''
specific features (e.g. frequencies) in the post-merger spectra
and employ EOS-insensitive relations (or quasi-universal relations, QUR) to
unveil EOS properties (e.g. phase transitions),
e.g.~\cite{Bauswein:2018bma,Breschi:2019srl,Weih:2019xvw,
	Prakash:2021wpz,Raithel:2022orm,Wijngaarden:2022sah}.
Only few authors have, however, considered the actual observational
and data analysis problem, namely, the problem of how to
incorporate these speculative ideas into a rigorous Bayesian data analysis
framework \cite{Breschi:2019srl,Wijngaarden:2022sah}. 
This paper discusses one
possible concrete method in this direction and some related conceptual
limitations in the realization of this program.

New degrees of freedom or phase transitions can impact the BNS
remnant dynamics at densities $\rho \gtrsim 2~\rho_{\rm sat}$, where
$\rho_{\rm sat}\simeq 2.7{\times}10^{}$ is nuclear saturation density,  
and leave signatures in the observable GWs. 
Case studies simulated BNSs with matter models including hyperon
production~\citep[e.g.][]{Sekiguchi:2011mc,Radice:2016rys,Bauswein:2018bma}
or zero-temperature models of phase transitions to quark-deconfined
matter~\citep[e.g.][]{Bauswein:2018bma,Most:2018eaw,Prakash:2021wpz,Fujimoto:2022xhv}.  
In these examples, a EOS softening with respect to the ``baseline''
hadronic EOS can determine a more compact remnant that either
undergoes an earlier gravitational collapse or increases the post-merger GW peak frequency $f_2$
towards higher values. The former case is particularly relevant for
binary masses above the prompt collapse threshold for the softened
EOS, but below that threshold for the hadronic EOS. This implies that one of
the two EOS model could be ruled out simply by the observation of a post-merger signal.
The latter case might instead be probed, in a suitable mass range, by
observing a violation (breakdown) of the QUR that relates $f_2$ to
properties of the individual neutron star (NS) in the binary,
e.g.~\cite{Bauswein:2018bma,Breschi:2019srl,Raithel:2022orm,Breschi:2022ens}. 
It is worth remarking that the detectability of these
effects crucially depends on the densities at which the EOS softening
takes place. Significant effects have been simulated by constructing
rather ``extreme'' transitions. 

EOS-insensitive relations are heavily used in 
GW astronomy with BNSs in order to either reduce the
matter's degrees of freedom in waveform modeling or connect spectral
features to the NS equilibria and mass-radius
diagram~\citep[e.g.][]{,Bernuzzi:2014kca,Yagi:2016bkt,Bauswein:2017aur,Godzieba:2020bbz}. 
Our work focuses on the relation between the dominant quadrupolar
spectral peak of the post-merger signal, $f_2$, and the
(leading order) tidal coupling constant $\kt$ of the
binary~\cite{Bernuzzi:2015rla,Breschi:2019srl}. 
This QUR allowed us to construct a unified full-spectrum model by combining an
inspiral-merger (IM) tidal waveform with a post-merger completion
\cite{Breschi:2019srl,Wijngaarden:2022sah,Breschi:2022ens,Breschi:2022xnc,Puecher:2022oiz}. 
Such relation represents a natural (and representative) choice for a pre/post-merger (PPM)
consistency test. 
To date, the employment of QURs is also the only method used in 
rigorous Bayesian studies, e.g.~\cite{Breschi:2019srl,Breschi:2021xrx,Wijngaarden:2022sah,
	Puecher:2022oiz}, to connect the binary properties to the post-merger features.

Inferring a QUR breakdown can be naturally treated as a PPM consistency test 
for a given QUR, similarly to analyses of
binary black hole (BBH) mergers in the context of tests of General
Relativity~\cite{Ghosh:2016qgn,Breschi:2019wki,LIGOScientific:2019fpa}.
We naturally employ such well-established framework to the analysis of BNS transients and demonstrate how to infer a QUR breakdown using 
Bayesian analyses of the full BNS spectrum.

The paper is structured as follows.
In Sec.~\ref{sec:methods}, we introduce the method used to detect
departures from quasi-universality. 
In Sec.~\ref{sec:results}, we validate our method performing parameter estimation (PE) on
mock GW data. 
Finally, we conclude in Sec.~\ref{sec:conclusions} highlighting
conceptual issues in the interpretation of the analysis in real GW
observations.

\section{Methods}
\label{sec:methods}

QUR breaking occurs when the quasi-universal prediction
does not match the corresponding observed property.
For the case of the post-merger peak $f_2(\kt)$,
the QUR is established as function of the binary properties,
that can be well-estimated from pre-merger GWs.
However, the post-merger signal directly provides a measurement of 
the $f_2$ frequency.
Thus, in order to identify the QUR breaking, we compare the post-merger
observations to the pre-merger predictions estimated with QURs.
Following the approach of Ref.~\cite{Ghosh:2016qgn},
we introduce a consistency test that aims to reveal such
breaking employing full-spectrum observations of BNSs.

Given the GW data and a waveform template, 
the posterior distributions of the BNS parameters are calculated via
Bayesian PE analysis~\citep[see, e.g.][]{Veitch:2009hd,Veitch:2014wba,Breschi:2021wzr}.
For our studies, we make use of the time-domain effective-one-body
(EOB) model {\teob}~\cite{Nagar:2018zoe} 
extended with the {\oldmodel} template in the high-frequency post-merger regime~\cite{Breschi:2019srl}.
In order to speed up the computations, 
the EOB template makes use of a reduced-order approximation~\cite{Lackey:2016krb}.
The considered post-merger model incorporates QURs calibrated on NR data,
used to predict the template features and it includes a
characterization of the main peaks of the post-merger spectrum.
Closely following \cite{Breschi:2019srl}, we perform three PE analyses:
first, we analyze the inspiral-merger data only (labeled as `IM')
with {\teob};
then, the post-merger data only (labeled as `PM')
is studied with {\oldmodel}, and,
finally,
we perform PE on the full-spectrum data (labeled as `IMPM')
with the complete model {\teob{\tt\_}\oldmodel}.

As discussed in Ref.~\cite{Ghosh:2016qgn},
PPM consistency tests relies on a 
cutoff frequency $f_{\rm cut}$ used to split the low-frequency and
high-frequency regimes. 
In general, the time-domain post-merger signal will also include frequency contributions below the merger frequency $f_{\rm mrg}$, due to the low quality factor of the QNMs dominating the remnant BH response.
However, for systems dominated by the quadrupolar mode, this ``mixing'' is typically negligible, and the portion of the signal with $f < f_{\rm mrg}$ only suffers from small contaminations from the time-domain post-merger phase. For this reason, choosing $f_{\rm cut} = f_{\rm mrg}$ is a sensible choice.
The ``mixing'' becomes more significant for lower remnant spins (induced e.g. by a nonspinning high mass ratio binary).
We stress that even in this case the consistency test remains valid, although the physical interpretation of the results becomes less immediate,
since a good fraction of a deviation in the $f < f_{\rm mrg}$ region could be induced by the time-domain post-merger signal.
For BNS signals, the post-merger signal can lead to significant spectral contamination
below $f_{\rm cut}$ and the split is less trivial.
However, if the dominant post-merger frequencies are significantly larger than
the merger frequency $f_{\rm mrg}$ or if the post-merger signal-to-noise-ratio (SNR)
contribution below the cutoff is negligible, one can still choose
$f_{\rm cut}= f_{\rm mrg}$.
This is the choice made in this work, assuming the cutoff frequency to be known exactly. 
In a realistic scenario, the cutoff frequency 
can be estimated from the full-spectrum posterior
using EOS-insensitive relations for the merger frequency for the
quadrupolar
mode~\cite{Breschi:2019srl,Bernuzzi:2020tgt,Breschi:2022xnc}.
If the splitting frequency $f_{\rm cut}$ cannot be uniquely fixed 
(e.g. due to spectral contamination below this threshold),
the `IM' and `PM' models might be treated
separately in single analyses either in a direct time-domain analysis~\cite{Carullo:2019flw,Isi:2021iql},
or augmenting the standard frequency domain likelihood using ``gating'' techniques~\cite{Zackay:2019kkv, Capano:2021etf, Isi:2021iql}. 
However, both of these methods are expected to significantly increase the computational cost,
compounding the already long computational times inherent in inspiral BNS analyses.

The `IM' inference provides direct information on the progenitors' properties
(i.e. masses, spins, tidal polarizabilities, \dots).
From these parameters, it is possible to estimate a prediction
for the $f_2$ posterior using the QUR in Eq.~13 of \cite{Breschi:2019srl}.
Also the `PM'  inference provides information on the the progenitors'
properties through the internally employed QURs. Moreover, in this
case,  the $f_2$ posterior can be directly estimated  
from the reconstructed waveform.
Finally, the `IMPM' case naturally delivers 
information on the progenitors' properties and it allows us to 
estimate the $f_2$ posterior from the reconstructed waveform.
Then, following the approach of Ref.~\cite{Ghosh:2016qgn},
we introduce  the (fractional) deviations from the QUR as
\begin{equation}
\label{eq:ng:df-f}
\frac{\Delta f_2}{f_2} = \frac{f_2^{\rm PM}-f_2^{\rm IM}}{f_2^{\rm IMPM}}\,,\quad 
\frac{\Delta \kt}{\kt} = \frac{{\kt}^{\rm PM}-{\kt}^{\rm IM}}{{\kt}^{\rm IMPM}}\,.
\end{equation}
We remark that $f_2^{\rm IM}$ is computed from the inspiral data using the QUR in post-processing,
while $f_2^{\rm PM}$ and $f_2^{\rm IMPM}$ estimation includes directly the PM data. 

The computation of $\pr({\Delta f_2}/{f_2}, \Delta \kt/\kt)$ 
is performed with a probabilistic approach.
Given the posteriors $\{f_2,\kt\}_i$ for $i={\rm IM}, {\rm PM}, {\rm IMPM}$,
the posterior of ${\Delta f_2}$ and 
${\Delta \kt}$ are estimated as
\begin{widetext}
\begin{equation}
\label{eq:ng:p-deltaf2-deltakt}
\pr(\Delta f_2,\Delta \kt|\data_{\rm IM},\data_{\rm PM}) = \iint \pr( f_2, \kt|\data_{\rm PM}) \,\pr(\kt-\Delta \kt, f_2-\Delta f_2|\data_{\rm IM})\,\d f_2 \,\d\kt\,. 
\end{equation}
\end{widetext}
Eq.~\eqref{eq:ng:p-deltaf2-deltakt} is the convolution product between 
the IM and the PM posteriors.
Then, 
labeling $\varepsilon_{f_2} = {\Delta f_2}/{f_2}$ and $\varepsilon_{\kt}= {\Delta \kt}/{\kt}$,
the posterior for 
the quantities in Eq.~\eqref{eq:ng:df-f} can be computed 
from the recovered posterior  as
\begin{widetext}
\begin{equation}
\label{eq:ng:p-deltaf2-deltakt-eps}
\pr(\varepsilon_{f_2},\varepsilon_{\kt}) = \iint \kt \, f_2 \, \pr( \varepsilon_{f_2} \cdot f_2, \varepsilon_{\kt}\cdot\kt|\data_{\rm IM},\data_{\rm PM})\, \pr(f_2, \kt|\data_{\rm IMPM})\,\d f_2 \,\d\kt\,. 
\end{equation}
\end{widetext}

As discussed in Ref.~\cite{Ghosh:2016qgn},
$\pr(f_2, \kt|\data_{\rm IMPM})$ represents our best guess for the $\{f_2, \kt\}$ posterior
and it is used in Eq.~\eqref{eq:ng:p-deltaf2-deltakt-eps} to weight the contributions
of the inspiral-merger and post-merger inferences;
while, $\pr(\Delta f_2,\Delta \kt|\data_{\rm IM},\data_{\rm PM})$ encodes 
the agreement/disagreement between pre-merger and post-merger inferences.
Within this approach the origin of the axes, i.e. ${\Delta f_2}=0$ and ${\Delta \kt}=0$,
represents the null-hypothesis for which no deviation from quasi-universality is observed.
On the other hand, a departure of the posterior  from the null-hypothesis
can indicate the breakdown of the $f_2(\kt)$ QUR.
Following the EOS terminology,
we label as a {\it softening} effect a deviation towards the region with 
${\Delta f_2}/{f_2}>0$ and ${\Delta \kt}/{\kt}<0$,
in order to differentiate it from a {\it stiffening} effect,
which shows ${\Delta f_2}/{f_2}<0$ and ${\Delta \kt}/{\kt}>0$.

\section{Results}
\label{sec:results}

\begin{figure}[t]
	\centering 
	\includegraphics[width=.49\textwidth]{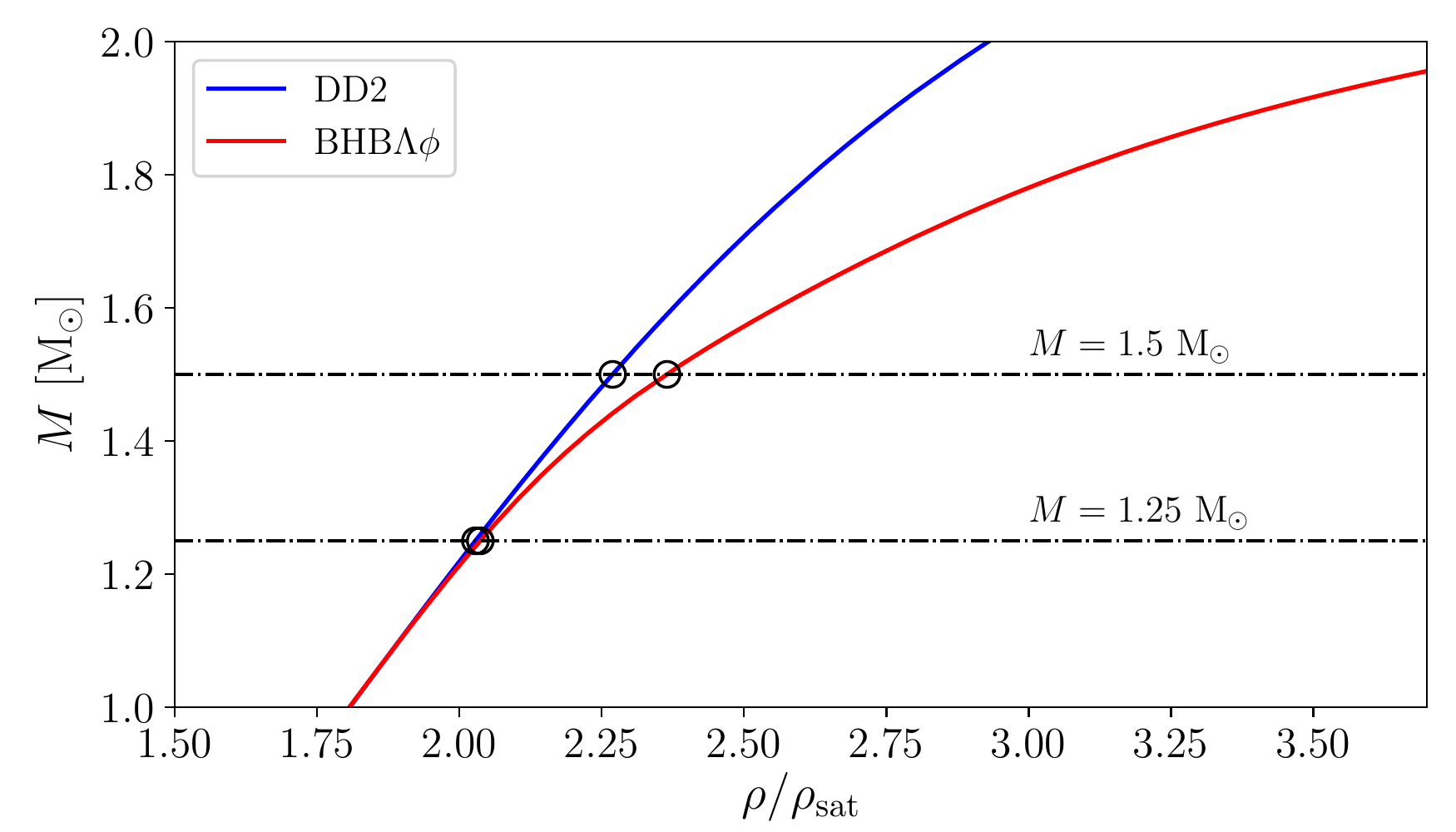}
	\includegraphics[width=.49\textwidth]{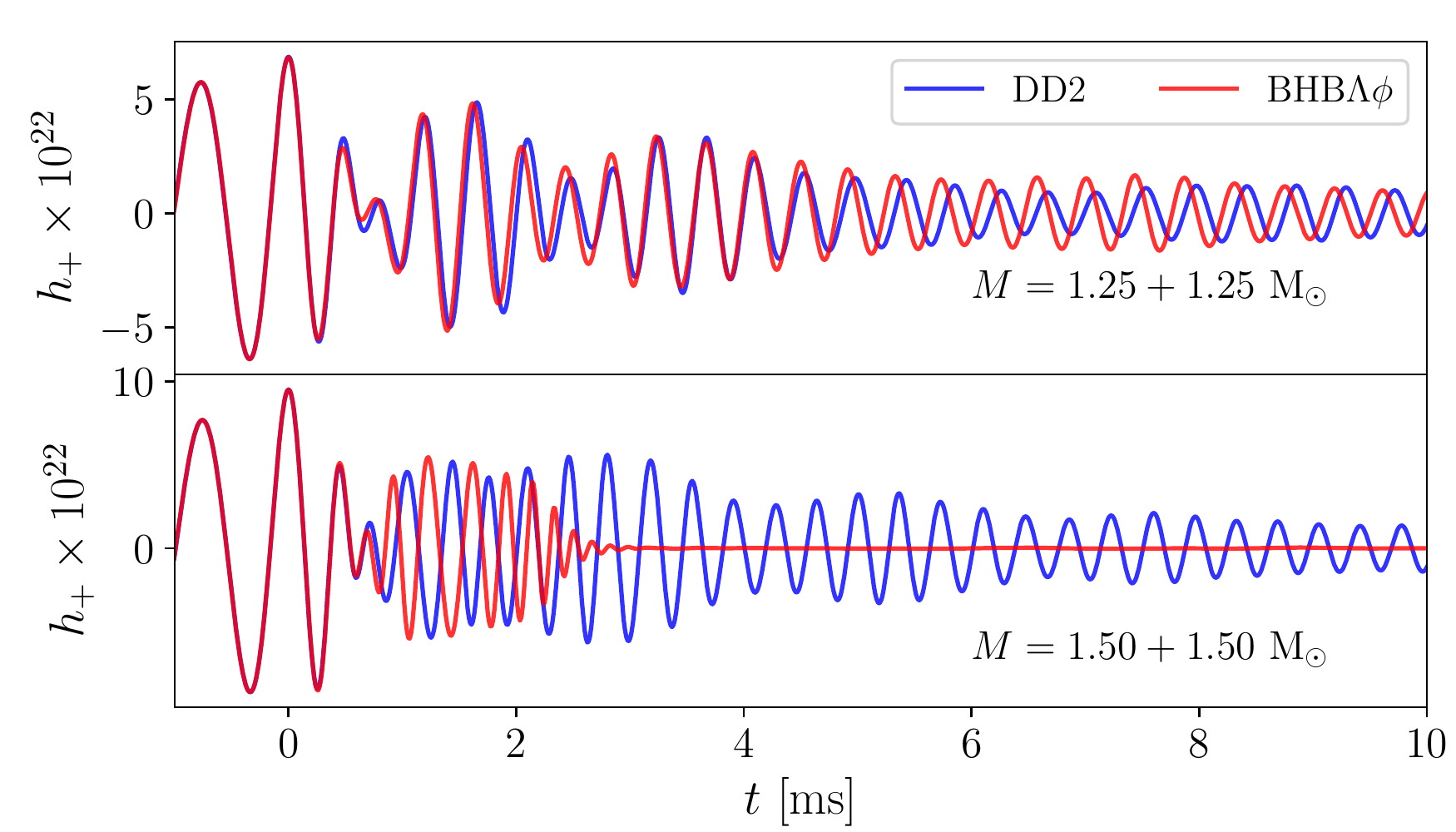}
	\caption[DD2 and BHB$\Lambda\phi$]{Comparison
		between the BHB$\Lambda\phi$ (red)
		and the DD2 (blue) EOS and 
		the corresponding BNS templates~\cite{Radice:2016rys}.
		{\it Top panel}: Mass of individual NSs as a function
		of the central density. The markers refer to simulated BNSs.
		{\it Bottom panel}: Plus polarization $h_+(t)$ of the NR waveforms 
		for the simulated BNSs with mass $M=2.5~\Mo$ (top) 
		and $M=3~\Mo$ (bottom).
		The binary are located at a fiducial distance of 40~Mpc.
		The origin of the time axis $t=0$ corresponds to the moment of merger.}
	\label{fig:ng:soft}
\end{figure}

We demonstrate the possibility of investigating the QUR breaking 
using PE analyses of mock GW data.
We discuss the specific case of
BHB$\Lambda\phi$ and DD2 EOS simulated in~\cite{Radice:2016rys}.
The BHB$\Lambda\phi$ EOS is identical to DD2 except that at densities
$\rho\gtrsim2.5\rhosat$ it softens due to the formation of $\Lambda$-hyperons.
Inspiral-merger GW signals from (equal-mass) binaries described by the two EOS and
$M\lesssim2.8~\Mo$ are indistinguishable since the
individual progenitor NSs have maximal densities $\rho\lesssim2.5\rhosat$, 
similar compactnesses and tidal parameters,
as shown in Figure~\ref{fig:ng:soft} (left).
On the other hand, for $M\gtrsim 2.8~\Mo$ the post-merger remnants reach higher
densities at which the two EOS differ, leading to different post-merger GWs 
as shown in Figure~\ref{fig:ng:soft} (right).

We consider a pair of high-mass binaries with $M=3~\Mo$, no spins and equal component masses
extracted from the {\tt CoRe} database~\cite{Dietrich:2018phi,Gonzalez:2022mgo}.
The individual progenitors of the high mass BNS have $\rho\approx2.35\rhosat$;
while, the associated remnant reaches $\rho\approx2.8\rhosat$ and
the presence of $\Lambda$-hyperons significantly affect the post-merger dynamics. 
The DD2 $1.50{+}1.50~\Mo$
binary has $f_2\simeq 2.76~{\rm kHz}$ and the respective
BHB$\Lambda\phi$ remnant has $f_2\simeq 3.29~{\rm kHz}$~\footnote{See Ref.~\cite{Breschi:2022xnc,Breschi:2022ens}
for discussions on the $f_2$ estimation for this case.}.
The difference
between the two NR values is ${\sim} 500~{\rm Hz}$, which corresponds
to ${\sim }20\%$. The BHB$\Lambda\phi$ data deviates of ${\sim}3{-}\sigma$
from  the prediction of the QUR presented in Ref.~\cite{Breschi:2019srl}
and employed in {\oldmodel}
($f_2^{\rm fit}=2.88~{\rm kHz}$), 
corresponding to a more compact remnant than the DD2 case. 
The two binaries have also different times of black-hole collapse: 
the DD2 case collapses at late times, i.e. ${\sim} 21~{\rm ms}$ after merger; 
while, the BHB$\Lambda\phi$ remnant collapses
shortly after merger within $2.6~{\rm ms}$.
Moreover, we repeat the analysis on the low-mass BHB$\Lambda\phi$ binary
with $M=2.5~\Mo$, whose morphology is almost identical to the corresponding DD2 
case even in the post-merger phase.
The corresponding waveforms are shown in Figure~\ref{fig:ng:soft} (right).
The data are generated as EOB-NR hybrid waveforms injected in zero-noise, 
while the recovery is performed using {\teob{\tt\_}\oldmodel}. 

We analyze 128~s of data with a lower frequency $f_{\rm low}=20 \rm Hz$ (or $f_{\rm low}= f_{\rm mrg}$ in the post-merger only case) and a sampling rate of 8192~Hz,
injecting the signal with post-merger SNR 11 (total SNR ${\sim}200$) and
using the three-detector LIGO-Virgo network at design sensitivity~\cite{Aasi:2013wya,
	TheVirgo:2014hva}.
The priors on the parameters are taken 
consistently with Ref.~\cite{Veitch:2009hd,Veitch:2014wba}
with spin parameters fixed to zero.
The PE studies are performed with the 
nested sampling routines implemented in {\tt LALInference}~\cite{Veitch:2009hd,Veitch:2014wba,lalsuite}
\footnote{The analysis settings are identical to Ref.~\cite{Breschi:2019srl}.
There, the reader can also find detailed 
discussion on the posteriors.}.

\begin{figure}[t]
	\centering 
	\includegraphics[width=0.49\textwidth]{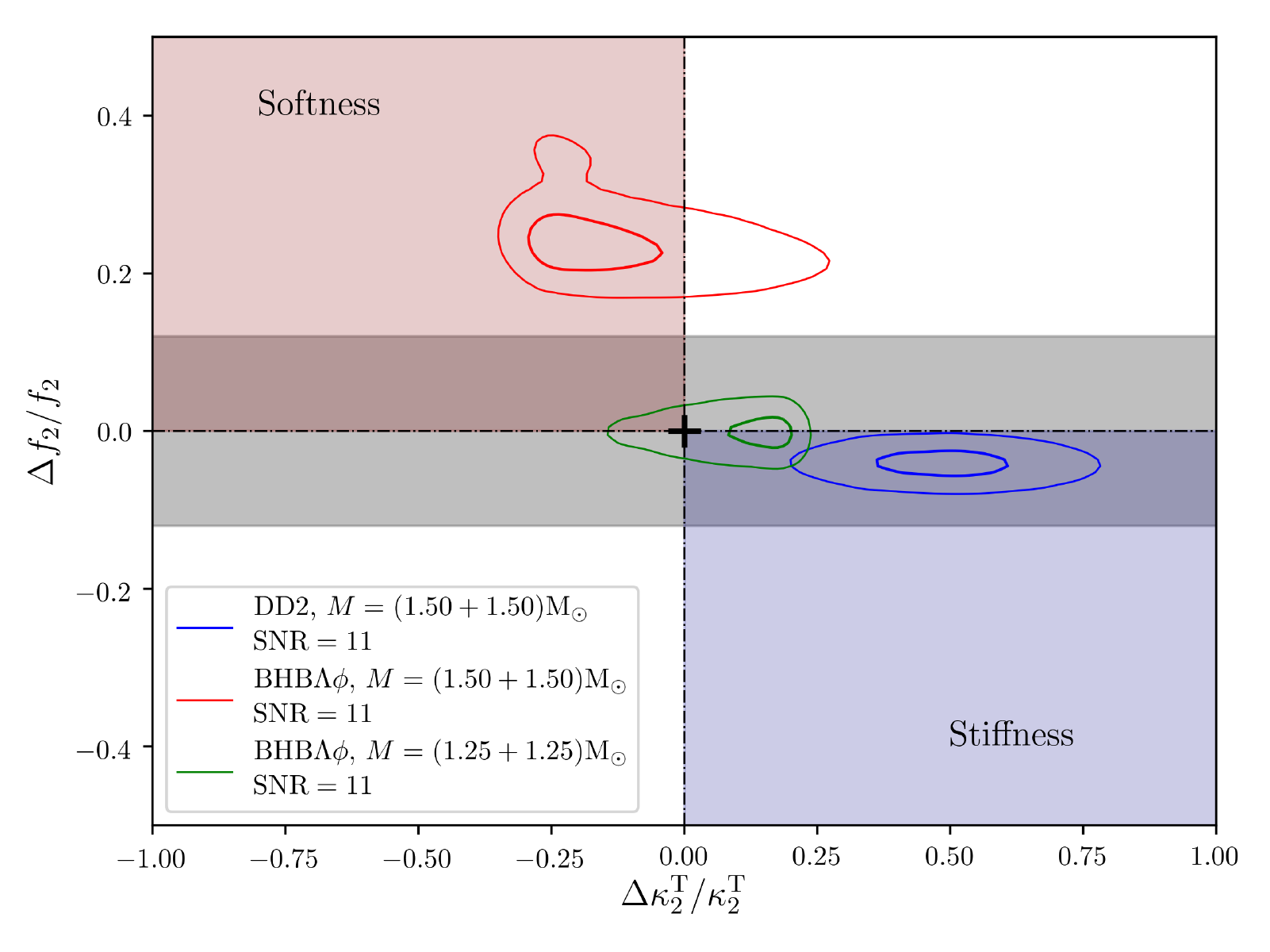}
	\caption[Consistency test for softening effects]{
		Posterior  for the deviation from the quasiuniversality defined
		in Eq.~\ref{eq:ng:df-f} for characteristic post-merger frequency $f_2$
		and tidal coupling $\kt$.
		The contours report the 50\% and the 90\% credibility regions.
		Red lines refer to low-mass BHB$\Lambda\phi$ binary,
		blue lines refer to high-mass DD2 binary 
		and red lines refer to high-mass BHB$\Lambda\phi$ binary.
		The red area denotes deviations due to softening effects, while blue 
		area identify the stiffening effects.
		The grey band report the 90\% credibility region of the $f_2$
		EOS-insensitive relation.
	}
	\label{fig:ng:soft_consist}
\end{figure}

Figure~\ref{fig:ng:soft_consist} shows the posterior 
estimated for the three considered binaries.
The grey band indicates the uncertainty of the QUR, and $\Delta f_2/f_2$ posteriors falling in this band are considered to be consistent with the assumed QUR.
The low-mass BHB$\Lambda\phi$ case
confidently includes the null hypothesis within the 90\% confidence level
of the posterior.
The $\Delta f_2/f_2$ posterior for the high-mass DD2 case is fully consistent with the QUR uncertainties,
indicating no significant deviation.
The mild deviation of $\Delta\kt/\kt=0.5^{+0.3}_{-0.3}$ toward the stiffness portion of the plane is due to the finite faithfulness of {\oldmodel} against the full NR simulation considered, and is expected to be cured by improved models~\cite{Breschi:2022xnc,Breschi:2022ens}.
The salient point to be exctraced from the figure, is that the high-mass BHB$\Lambda\phi$ case shows a significant
deviations toward the softness portion of the plane, with 
$\Delta \kt/\kt=-0.2^{+0.5}_{-0.2}$
and 
$\Delta f_2/f_2=0.2^{+0.2}_{-0.1}$.
This deviation in the frequency is significantly above the fit uncertainty
and demonstrates a successful detection of the QUR breaking, 
invalidate the applicability of the QUR $f_2(\kt)$ to the
considered binary.

\section{Conclusions}
\label{sec:conclusions}

Our results demonstrates a quantitative Bayesian method to invalidate a
given QUR using full-spectrum BNS observations. 
The observation of an inconsistency in a PPM 
analysis of this type might help to {\it exclude}
(some of) the EOS employed for the design of the QUR.
Although in the specific case considered this inconsistency was indeed caused by the appearance 
of hyperons at high densities (a ``phase transition''), we stress that demonstrating the breakdown of a QUR within a
given confidence level does not necessarily imply the measurement
of a EOS softening effect.
Since the true EOS is not known, but the inference requires a model 
(the QUR) designed using a EOS sample, it is only possible to
invalidate the model (hypothesis) using the proposed null test.
For example, this consistency test might simply exclude a QUR which is
``not sufficiently'' EOS-insensitive or which is poorly
designed. Ref.~\cite{Breschi:2022xnc} discusses the specific case of
$f_2(R_{1.4})$, where $R_{1.4}$ is the radius of an equilibrium NS of
mass $1.4M_\odot$. According to current available data and EOS models,
the $f_2(R_{1.4})$ QUR might be easily broken by an observation at minimal post-merger SNR
for detection. However, if one considers a similar QUR with the same
quantities but rescaled by the binary mass, the QUR significantly
improves its EOS-insensitive character.
We stress that, according to current theoretical models and
constraints, demonstrating the breaking of a (well-designed) QUR requires significant fine-tuning of both
the EOS model and the binary masses, Cf.~\cite{Radice:2016rys,Bauswein:2018bma,Prakash:2021wpz,Wijngaarden:2022sah}.

The presented method is not restricted to the particular QUR considered
here. A similar analysis may be performed, for example, on the
inferred collapse time \cite{Breschi:2022ens}, considering the
consistency of multiple parameters/QUR involved in the GW template, or
using other
QURs~\cite[e.g.][]{Bauswein:2018bma,Raithel:2022orm}.  
However, the $f_2(\kt)$ QUR is particularly interesting because (i) it
is directly involved in the construction of the GW template, and (ii)
it is rather accurate and shows deviations at a few percent level
although being built from the largest sample of EOS and simulations
explored so far in numerical relativity.
Improved analyses can be obtained by folding-in recalibration
parameters to better account for the uncertainties of the QUR, as
shown in Ref.~\cite{Breschi:2021xrx,Breschi:2022ens,Breschi:2022xnc}. 

BNS post-merger signals are likely to be accessible with
next-generation ground-based GW
interferometers for events comparable (or louder) than
GW170817~\citep[e.g.][]{Punturo:2010zza,Hild:2011np,Breschi:2022ens}.  
In order to gain information on the nuclear matter from these
observations, it seems necessary to significantly extend current
theoretical EOS models and simulations and explore within Bayesian
analysis frameworks such predictions.

\section*{Acknowledgments}

MB and SB acknowledge 
support from the European Union’s H2020 ERC Starting
Grant, no.~BinGraSp-714626.
MB acknowledges support from the Deutsche Forschungsgemeinschaft
(DFG) under Grant no.~406116891 within the Research Training Group
(RTG) 2522/1.
MB acknowledges support from the European Union’s H2020 ERC
Consolidator Grant “GRavity from Astrophysical to Microscopic Scales”
(Grant no.~GRAMS~815673) and the EU Horizon 2020 Research and
Innovation Programme under the Marie Sklodowska-Curie Grant Agreement
No.~101007855.
GC acknowledges support by the Della Riccia Foundation under an Early Career Scientist Fellowship.
GC acknowledges funding from the European Union’s Horizon 2020 research and innovation program under the Marie Sklodowska-Curie grant agreement No. 847523 ‘INTERACTIONS’, from the Villum Investigator program supported by VILLUM FONDEN (grant no.~37766) and the DNRF Chair, by the Danish Research Foundation.
SB acknowledges support from the DFG project MEMI no. BE 6301/2-1.
The computational experiments were performed on the {\scshape Tullio}
sever at INFN Turin. 
The waveform model employed in this work, {\teob{\tt \_}\oldmodel}, 
is implemented in {\scshape bajes} and the software is publicly available at:

\url{https://github.com/matteobreschi/bajes} 


%


\end{document}